\documentclass[11pt,a4paper]{article}
\usepackage[intlimits,centertags]{amsmath}
\usepackage{amssymb,amsfonts}
\usepackage{bbm}
\usepackage{mathrsfs}
\usepackage{cite}
\usepackage[scale={0.75,0.81}]{geometry}
\usepackage{multirow}
\usepackage{relsize}
\usepackage[subrefformat=parens,labelformat=parens]{subfig}
\usepackage{color}
\usepackage{caption}
\usepackage{graphics,graphicx}
\newlength{\textlength}
\newlength{\overlinelength}
\newcommand{\ol}[2][.625]{%
   \settowidth{\textlength}{\ensuremath{#2}}%
   \setlength{\overlinelength}{3pt}%
   \addtolength{\overlinelength}{0.4\textlength}%
   \makebox[\textlength][s]{\ensuremath{#2}}%
   \hspace{-.5\textlength}\hspace{-\overlinelength}\hspace{#1\overlinelength}
   \overline{%
      \makebox[\overlinelength][c]{%
         \vphantom{\ensuremath{#2}}
      }
   }
   \hspace{-#1\overlinelength}\hspace{.5\textlength}
}

\newcommand{\I}{\ensuremath{\text{i}}}
\newcommand{\D}{\ensuremath{\text{d}}}

\DeclareMathOperator{\tr}{tr}

\allowdisplaybreaks[2]
\numberwithin{equation}{section}

\begin{document}
  \title{Non-Universal Anomalies in Heterotic String Constructions}
  \author{Christoph L\"{u}deling, Fabian Ruehle, Clemens Wieck\\[2mm]
    {\normalsize\itshape Bethe Center for Theoretical Physics and Physikalisches Institut der
      Universit\"at Bonn,}\\{\normalsize\itshape Bonn, Germany}\\ 
    {\small\ttfamily luedeling, ruehle, wieck @th.physik.uni-bonn.de}}
  \date{}

\maketitle

\begin{abstract}
We investigate anomalies on heterotic orbifolds and their blowups. We give a simple example of an orbifold blowup which contains anomalous $U(1)$ symmetries that are canceled by axions which couple non-universally to the different gauge groups, thus clarifying some confusion which recently arose in the literature concerning anomaly universality. We argue that non-universal axionic couplings are the general case, and that the couplings are only universal in the case of orbifolds. We comment on the consequences of this non-universality for discrete $R$ symmetries. We furthermore investigate the origin of discrete ($R$ and non-$R$) symmetries on smooth Calabi--Yau manifolds. 
\end{abstract}

\section{Introduction} 
\label{sec:Introduction}
String theory is probably the best-developed theory to describe a UV completion of the Standard Model. It has therefore proven very important in the study of physics beyond the Standard Model. On top of providing a UV completion of the theory, it has an elegant way of yielding anomaly-free and thus consistent theories. Constructing anomaly-free theories is a formidable task, especially in higher dimensions, and if it was not for string theory, probably far fewer consistent constructions would be known.

However, anomaly freedom in string theory still imposes strong constraints on the gauge groups. For heterotic string theory \cite{Gross:1984dd,Gross:1985fr}, which will be dealt with in this paper, the only two possible gauge groups are $SO(32)$ and $E_8\times E_8$. We will only consider the latter in this paper, but the results are of course also valid in other contexts. In addition to constraining the possible gauge groups, anomaly cancellation requires the presence of axions. The anomaly cancellation mechanism via axions is known as the Green--Schwarz mechanism \cite{Green:1984sg}.

In order to describe a heterotic string theory, one has to specify an underlying compactification geometry and a compatible description of the gauge sector. In this paper, we will deal with 4d theories obtained from heterotic orbifold constructions \cite{Dixon:1985jw,Dixon:1986jc,Ibanez:1986tp} and their smooth Calabi--Yau counterparts which are obtained via a blowup procedure (see \cite{Candelas:1985en,Aspinwall:1994ev,Nibbelink:2007rd,Nibbelink:2007pn} and \cite{NibbelinkGroot:2010wm,Blaszczyk:2011hs} for a gauged linear sigma model (GLSM) approach). 

In orbifold constructions, imposing modular invariance of the string partition function is sufficient to guarantee consistency of the theory via the Green--Schwarz mechanism \cite{Schellekens:1986xh}, where the axion dual to the Kalb--Ramond $B_2$ field cancels the anomalies. Since there is only one axion but many different possible anomalies, anomaly freedom requires that all anomalies are related such that the different anomalies can be canceled with one universal axion. This requires the presence of at most one anomalous $U(1)$ as well as a universal coupling of the axion to the spin and gauge connection.

For Calabi--Yau compactifications, the picture is different. While the Bianchi identity ensures the absence of purely non-Abelian anomalies, mixed Abelian--non-Abelian anomalies can still arise. These are again canceled via the Green--Schwarz mechanism \cite{Blumenhagen:2005ga,GrootNibbelink:2007ew,Blaszczyk:2011ig,Lukas:1999nh}. However, in contrast to the orbifold case, there can be more than one axion, and the axionic couplings are in general non-universal, even if there is only one axion. The reason is that in blowup, the additional axions arise from internal cohomology 2-forms and their couplings are determined by the choice of the gauge bundle. This anomaly cancellation mechanism becomes important when considering line bundles over Calabi--Yau spaces (see e.g.\ \cite{Blaszczyk:2010db,Anderson:2011ns,Anderson:2012yf} for recent applications).

Both the orbifold and the Calabi--Yau constructions are in general plagued by the common problems of theories
beyond the Standard Model, like too fast proton decay and the $\mu$ problem. One of the prevailing methods of
circumventing these problems are discrete symmetries which forbid the associated superpotential terms. For
orbifold constructions, it has recently been argued \cite{Lee:2010gv} that there is a unique $\mathbbm{Z}_4^R$
symmetry which fulfills all constraints and is compatible with $SO(10)$ grand unification. The authors used
anomaly universality in their uniqueness argumentation. While correct for pure orbifold constructions, anomaly
universality is in general not satisfied, and one purpose of this paper is to clarify this confusion: As we
discuss, universality is neither required from a bottom-up point of view nor expected for general heterotic
Calabi--Yau constructions.

In order to illustrate these points, we present a rather simple model, which is based on the blowup of the $T^6/\mathbbm{Z}_3$ orbifold with line bundles, such that only two $U(1)$ factors arise. The anomalies are non-universal and canceled by the non-universal axions from the reduction along the internal 2-forms. Along the way, one of the models also provides a simple example for a $U(1)$ which is omalous\footnote{We use the term omalous coined by Donagi for non-anomalous symmetries.} but nevertheless massive. In addition, we use the model to illustrate how discrete ($R$ and non-$R$) symmetries arise in Calabi--Yau models. Ultimately, these symmetries can be of utter importance for string phenomenology.

The paper is organized as follows: In Section~\ref{sec:GSMechanism}, we review anomalies and the Green--Schwarz mechanism. In Section~\ref{sec:StringModel}, we discuss the consequences of non-universal anomalies. First we discuss a bottom-up approach and argue that anomaly universality is generically not realized. Then, we discuss a top-down approach using a simple string realization of non-universal anomaly cancellation. We present our model and work out the anomalies and the masses of the Abelian gauge bosons. In Section~\ref{sec:DiscreteSymmetries}, we investigate the presence of discrete $R$ and non-$R$ symmetries on the orbifold and on the resolved Calabi--Yau. Finally, in Section~\ref{sec:Conclusion} we conclude and give an outlook on future research directions.

\section{Anomaly cancellation and universality} 
\label{sec:GSMechanism}

In this section we will briefly review anomalies and the Green--Schwarz mechanism\cite{Green:1984sg} (for a review of anomaly cancellation in higher dimensions, see e.g.~\cite{Scrucca:2004jn}), with particular emphasis on ten- and four-dimensional supergravities derived from the heterotic string.

Anomalies arise if a classical symmetry of a theory is not preserved under quantization. This only happens for chiral fermions (and potentially other chiral fields, such as (anti)self-dual tensors, but these will not play a role in our discussion), and thus only in even dimensions $d$. If the anomalous symmetry is a gauge symmetry, the theory is actually inconsistent. 

For concreteness, consider a theory with a gauge group which is a product of a non-Abelian group $G$ (where $G$ can contain several factors) and a number of $U(1)$'s, and a set of massless chiral fermions~$f$, transforming in representations $\boldsymbol{r}_f$ of $G$ and with $U(1)$ charges $q_f^i$. In a Feynman diagram approach, anomalies arise via $(d/2+1)$-sided polygon graphs, with chiral fields in the  loop. For example, in four dimensions the relevant diagram is a triangle, and we can take all fields to be left-handed. Concentrating on the $U(1)$ anomalies, we can attach either three $U(1)$ gauge fields to the vertices, or one $U(1)$ and two non-Abelian gauge fields, including gravitons (for the purpose of anomalies, we can basically treat the Lorentz group as another gauge group factor). The diagram leads to an anomalous divergence of the $U(1)$ current $J_i$, 
\begin{align}
  \partial_\mu J_i^\mu&\sim A_{G^2-U(1)_i} \tr F_{\mu\nu}\widetilde{F}^{\mu\nu} + \frac{1}{s_{ijk}} A_{U(1)^3_{ijk}} F_{j\,\mu\nu}\widetilde{F}_k^{\mu\nu} +\frac{1}{24} A_{\text{grav}^2-U(1)_i} \tr R_{\mu\nu}\widetilde{R}^{\mu\nu} \,.
\end{align}
Here $F$, $F_i$ and $R$ are the field strengths of $G$, $U(1)_i$ and the Riemann tensor, the tilde denotes the dual (i.e.\ $\widetilde{F}^{\mu\nu}=\frac{1}{2} \varepsilon^{\mu\nu\rho\sigma}F_{\rho\sigma}$), and the traces are taken in a suitable representation.  Finally, $s_{ijk}$ is a symmetry factor taking into account permutations of the legs (i.e.\ for distinct $i$, $j$ and $k$, we have $s_{iii}=3!$, $s_{iij}=2!$ and $s_{ijk}=1$). The anomaly coefficients for these cases are given by 
\begin{align}
  \label{eq:anomcoeffs}
  A_{G^2-U(1)}&=\sum_{f} q_f \ell\!\left(\boldsymbol{r}_f\right)\,, & A_{\text{grav}^2-U(1)} &=\sum_m q_m\,,
  & A_{U(1)^3_{ijk}}&=\sum_m q_m^i q_m^j q_m^k\,.
\end{align}
Here the first sum runs over all chiral fermions $f$ transforming in the representation $\boldsymbol{r}_f$ of G, and $\ell\!\left(\boldsymbol{r}_f\right)$ is the quadratic index of $\boldsymbol{r}_f$. Similarly, the second and third sums run over all chiral fermions. 

These coefficients have the same form for $\mathbbm Z_N$ anomalies, but they only have to vanish mod $N$ (or mod $\frac{N}{2}$ if $N$ is even) to ensure consistency of the theory. Furthermore, it can be argued that cubic discrete anomalies are not relevant for the discussion \cite{Banks:1991xj}. When considering $R$ symmetries, the charges in \eqref{eq:anomcoeffs} have to be shifted by $-1$, since fermions run in the loop. Additionally, for $R$ symmetries the gauginos in the loop contribute a factor $\ell(\text{adj}_f)$ to \eqref{eq:anomcoeffs}.

\medskip

A conceptually clearer approach to anomalies and their cancellation is via the anomaly polynomial. Here an anomaly manifests itself as non-invariance of the effective action $\Gamma$ under a gauge (or local Lorentz) transformation with parameter $\lambda$, 
\begin{align}
  \label{eq:anomalousvarition}
  \Gamma&\longrightarrow \Gamma + \mathcal{A}\!\left(\lambda\right)\,.
\end{align}
Since the classical action is invariant, $\mathcal{A}$ arises from the path integral measure and can be calculated e.g.\ by Fujikawa's method \cite{Fujikawa:1979ay,Fujikawa:1980eg} or as the index of a suitable Dirac operator\cite{AlvarezGaume:1984dr}. It is given by an integral of a $d$-form, $\mathcal{A}=\int I_d^{(1)}$, and the Wess--Zumino consistency conditions imply that the nontrivial information about the anomaly can be succinctly given in terms of a (formal) closed and gauge-invariant $d+2$-form $I_{d+2}$, which is related to $I_d$ via the Stora--Zumino descent equations
\begin{align}
  \label{eq:descenteqns}
  I_{d+2}= \D I_{d+1}^{(0)}\,, \qquad \delta_\lambda I_{d+1}^{(0)}=\D I_d^{(1)}\,.
\end{align}
The anomaly polynomial $I_{d+2}$ is composed out of traces of powers of field strengths, including the Riemann tensor, treated as matrix-valued two-forms. The $I_{d+1}^{(0)}$ are called Chern--Simons forms. 

Generically, if the anomaly is nonzero, the theory is inconsistent. However, if the anomalies are reducible, they can be canceled by the Green--Schwarz mechanism. Reducible means that the anomaly polynomial factorizes, i.e.\ it can be written as a sum 
\begin{align}\label{eq:reducibleanomaly}
  I_{d+2}&=\sum_{a=1}^m X_{d+2-k_a} Y_{k_a}\,,
\end{align}
where each of the factors is by itself closed, gauge invariant and of even degree. (Throughout this paper, the wedge product for forms is understood.) This in particular requires that terms such as $\tr F^{d/2+1}$ are absent for gauge groups which have a Casimir invariant of that order. The idea of the Green--Schwarz mechanism is to compensate the variation of the effective action by an explicitly non-gauge invariant piece involving fields that transform with a shift. In a bottom-up approach, one may add these fields by hand, while in a top-down approach, such as string compactifications, these fields have to be present from the start.

Concretely, a reducible anomaly of the form~(\ref{eq:reducibleanomaly}) requires a set of $(k_a-2)$-form fields $C_{k_a-2}$, whose gauge transformation is a shift proportional to the descent of $Y_{k_a}$,  
\begin{align}
  \delta C_{k_a-2}&=-\xi Y_{k_a-2}^{(1)}\,.
\end{align}
Here $Y_{k_a-2}^{(1)}$ is the descent of the factor $Y_{k_a}$ in the anomaly polynomial, and $\xi$ is a free parameter. This transformation implies that the field strength of $C_{k_a-2}$ contains the associated Chern--Simons form, 
\begin{align}\label{eq:H=db+wCS}
  H_{k_a-1}&=\D C_{k_a-2}+ \xi Y_{k_a-1}^{(0)}\,,
\end{align}
and consequently the Bianchi identity for $H_{k_a-1}$ becomes
\begin{align}
  \D H_{k_a-1}&=\xi Y_{k_a}\,.
\end{align}
The anomalous variation~(\ref{eq:anomalousvarition}) of the action is now canceled by the Green--Schwarz action for the form fields\footnote{Note that there is in principle a third term $\sim Y_{k_a-1}^{(0)} X_{d+1-k_a}^{(0)}$. The ambiguity arises because the anomaly $I_d^{(1)}$ can be a linear combination of the descents along $X_{d+2-k_a}$ and $Y_{k_a}$. We have made a particular choice here for simplicity.},
\begin{align}
  \label{eq:GSaction}
  S_\text{GS}&=\sum_{a=1}^m \int \frac{1}{2} H_{k_a-1}\wedge *H_{k_a-1} +\frac{1}{\xi} C_{k_a-2} X_{d+2-k_a}\,.
\end{align}
Hence, each form field couples to some combination of gauge and gravitational field strengths encoded in $X_{d+2-k_a}$. It is, however, not required that the gauge group factors appear with universal coefficients in the coupling.

We have phrased the previous discussion in terms of the $C_{k_a-2}$ with $\delta C_{k_a-2}=-\xi Y_{k_a}^{(1)}$. We could just as well have described the mechanism in terms of the dual forms $\widetilde{C}_{d-k_a}$, for which one has to switch the roles of $X$ and $Y$. In particular, this dualization trades a coupling $C_{k_a-2}\wedge X_{d+2-k_a}$ for a shift $\delta\widetilde{C}_{d-k_a}=-X_{d-k_a}^{(1)}$.

\medskip

Let us now specialize to the case of interest, which are low-energy effective theories derived from the $E_8\times E_8$ heterotic string. Here we have exactly one field which can take part in the Green--Schwarz mechanism, namely the Kalb--Ramond two-form $B_2$, whose gauge transformation is fixed from string theory.

In ten dimensions, the low-energy theory is supergravity coupled to super-Yang--Mills theory. The anomaly polynomial $I_{12}$ receives contributions from the gravitino, the 
dilatino and  the gauginos, and the requirement of anomaly cancellation singles out $SO(32)$ and $E_8\times E_8$ as the only consistent gauge groups. (Upon replacing one or both $E_8$'s by $U(1)^{248}$, $I_{12}$ still factorizes, but the anomalies are not canceled as the $U(1)$'s appear in the transformation of $B_2$ but not in the anomaly polynomial\cite{Adams:2010zy}.)

The anomaly polynomial factorizes as 
\begin{align}
  \begin{split}
    \label{eq:anomalypolynomial10d}
    I_{12}&= Y_8 X_4=\left(\frac{1}{8}\tr \mathfrak{R}^4 +\frac{1}{32}\left(\tr\mathfrak{R}^2\right)^2-\frac{1}{8}\tr\mathfrak{R}^2 \tr \mathfrak{F}^2 +\frac{1}{24} \tr\mathfrak{F}^4-\frac{1}{8}\left(\tr \mathfrak{F}^2\right)^2\right)\\ 
    &\quad \mspace{80mu}\times \left(\tr \mathfrak{R}^2 -\tr \mathfrak{F}^2\right) \,.
  \end{split}
\end{align}
Here $\mathfrak{R}$ and $\mathfrak{F}$ denote the ten-dimensional Riemann and field strength tensors. For the $E_8\times E_8$ theory, $\mathfrak{F}=\mathfrak{F}'\oplus \mathfrak{F}''$, and the corresponding traces are sums of the terms for the two $E_8$'s. Explicitly, the two-form $B_2$
transforms as
\begin{align}
  \label{eq:deltaB2}
  \delta B_2&= \tr \Theta \D \Omega - \tr \lambda'\D \mathfrak{A}' -\tr \lambda'' \D \mathfrak{A}''\,,
\end{align}
where $\Omega$, $\mathfrak{A}'$ and $\mathfrak{A}''$ are spin and gauge connections and $\Theta$, $\lambda'$ and $\lambda''$ are the corresponding transformation parameters. The associated Chern--Simons three-forms are 
\begin{align}
  \omega'_3&= \tr \mathfrak{A}'\D \mathfrak{A}'-\frac{2\I}{3} \mathfrak{A}'^3 
\end{align}
and similar for the other connections. 

\medskip

For four-dimensional theories, factorizable anomalies can only take the form $I_6=\sum_a Y_2^a X^a_4$, and can thus be canceled by two-forms or scalars (axions). Since in four dimensions, these are dual to each other (in the sense that their field strengths satisfy $*H_3=H_1$), we can phrase the following discussion in terms of scalars only. The two-form $Y_2$ can only be a field strength of a $U(1)$ gauge group factor, $Y_2=\D A_1$. If such a factor appears, we call the $U(1)$ anomalous, while $U(1)$ factors which do not appear in this way are called omalous. Then $X_4$ involves the anomalous $U(1)$ and either two more $U(1)$'s or a square of a non-Abelian group (again including gravity), and by an abuse of notation we denote these as Abelian, non-Abelian and gravitational anomalies, respectively.

If the axion $C_0$ cancels the anomaly, its field strength~(\ref{eq:H=db+wCS}) becomes 
\begin{align}
  \label{eq:D(axion)}
  H_1=\D C_0+\xi A_1\,,
\end{align}
and thus the kinetic term for $C_0$ leads to a mass term $\sim \xi^2 \left|A_1\right|^2$. Thus, an anomalous $U(1)$ gets a St\"uckelberg mass from the Green--Schwarz axion, which can be gauged away and ``eaten'' by $A_1$.

\medskip

For heterotic $d=4$ models, we investigate two cases: Compactifications on orbifolds and on smooth
Calabi--Yaus with vector bundles, including orbifold blowups\footnote{For the question of localized anomalies
  on orbifold fixed points, see e.g.~\cite{Gmeiner:2002es,Buchmuller:2007qf}.}. In the first case, the only
field able to cancel anomalies is the four-dimensional two-form $b_2$ coming from the reduction of $B_2$,
while neither twisted nor untwisted sectors contain fields that transform with a shift. 
Hence $I_6$ factorizes into a single product,
$I_6=X_4 Y_2$. Furthermore, $X_4$ is simply given by the second line of Eq.~(\ref{eq:anomalypolynomial10d}),
restricted to four-dimensional forms and the unbroken gauge group. Upon dualizing $b_2$ to an axion $a$, the
anomaly is canceled by the coupling 
\begin{align}\label{eq:universalaxioncoupling}
  \int a \, X_4\,,
\end{align}
so in particular $a$ couples universally to all gauge groups. Note that this coupling arises in the dualization as a consequence of the gauge transformation properties of $b_2$, so it is actually independent of the existence of an anomalous $U(1)$, which manifests itself as a shift of $a$ under the anomalous symmetry. Furthermore, if one $E_8$ is unbroken, there will be no matter states charged under this $E_8$ and consequently no mixed anomaly to cancel. Hence the coupling~(\ref{eq:universalaxioncoupling}) cannot produce a gauge variation, and there is no anomalous $U(1)$ in this case. 

In the second case, there will generically be many additional axions $\beta_r$ arising from the reduction of $B_2$ along internal cohomology two-forms $E_r$. Their transformation follows from Eq.~(\ref{eq:deltaB2}) by expanding the internal flux and comparing the terms proportional to $E_r$. Hence these axions will not couple universally to all gauge groups (otherwise one could redefine them by a term proportional to the universal axion). So in this setup, there can be up to~16 anomalous $U(1)$'s, at most one of which couples universally. Note further that if an axion shifts under a $U(1)$, the gauge boson is massive by its field strength~(\ref{eq:D(axion)}) even if the $U(1)$ is omalous.

In particular, for a compactification on a Calabi--Yau $X$ with line bundles, the anomaly polynomial can be evaluated easily: Split the 10d gauge fields into internal background flux and four-dimensional fluctuations, $\mathfrak{F}=\mathcal{F}+F$, $\mathfrak{R}=\mathcal{R}+R$. Here we assume that the four-dimensional backgrounds vanish and that there are no massless internal fluctuations. The backgrounds satisfy the Bianchi identity 
\begin{align}
  \label{eq:dH-BI}
  \D H&=\tr \mathcal{R}^2 - \tr \mathcal{F'}^2 - \tr \mathcal{F''}^2 = 0
\end{align}
in cohomology, i.e.\ $\tr \mathcal{R}^2 - \tr \mathcal{F'}^2 - \tr \mathcal{F''}^2$ is an exact form that vanishes when integrated over any closed four-cycle. Then one can straightforwardly insert this split into the ten-dimensional anomaly polynomial\footnote{Since there are no purely gravitational anomalies in four dimensions, one can restrict to the gaugino contributions, as the gravitino and dilatino anomalies only involve the Riemann tensor.}~(\ref{eq:anomalypolynomial10d}) and keep the terms cubic in the backgrounds to get~\cite{Nibbelink:2009sp,Blumenhagen:2005ga}
\begin{align}
  \begin{split}
  \label{eq:4dpolynomial}
    I_6 &= \frac{1}{(2 \pi)^6} \mathlarger{\int}_X \left[ \frac{1}{6} \tr\!\left(\mathcal{F}' F'\right)^2 + \frac{1}{4} \left(\tr \mathcal{F}'^2 - \frac{1}{2} \tr\mathcal{R}^2 \right)\tr F'^2\right.\\
    &\quad\left.\mspace{70mu}- \frac{1}{8} \left(\tr \mathcal{F}'^2 - \frac{5}{12} \tr \mathcal{R}^2 \right)\tr R^2 \right] \tr\!\left(\mathcal{F}' F'\right) + \left(F',\mathcal{F}' \leftrightarrow F'',\mathcal{F}''\right)
  \end{split}
\end{align}
Each $E_8$ contributes three terms:
\begin{itemize}
  \item $\int_X\tr\!\left(\mathcal F' F'\right)^2 \cdot\tr\!\left(\mathcal{F}' F'\right)$ gives rise to Abelian anomalies only. This is true also for bundles with non-Abelian structure group $H$ because the generators of $H$, for which the trace gives a nonvanishing contribution, are broken by the bundle.
  \item $\int_X \left(\tr \mathcal F'^2 - \frac{1}{2} \tr \mathcal R^2 \right) \tr F'^2\cdot\tr\!\left(\mathcal F' F'\right)$  gives Abelian and non-Abelian anomalies, and
  \item $\int_X\left(\tr \mathcal F'^2 - \frac{5}{12} \tr \mathcal R^2 \right)\tr R^2\cdot \tr\!\left(\mathcal F' F'\right)$ leads to gravitational anomalies.
\end{itemize}
So we see that there is some partial anomaly universality: The non-Abelian anomalies coming out of the first $E_8$ are captured by one anomalous $U(1)$ factor with universal coefficients, and similar for the second $E_8$. Furthermore, if one $E_8$ is unbroken, i.e.\ $\mathcal{F}''=0$, the Bianchi identity~(\ref{eq:dH-BI}) implies that the non-Abelian and gravitational anomalies are captured by the same $U(1)$, and their coefficients are proportional to each other.

\section{Models with non-universal anomalies} 
\label{sec:StringModel}

\subsection{Bottom-up model}
\label{sec:BottomUpModel}

As a simple example of a model which does not require anomaly universality, consider an extension of the MSSM  by an additional $U(1)_X$. Here, the anomaly coefficients depend on the $U(1)_X$ charges of the MSSM superfields, i.e.\ on the  phenomenological requirements one imposes. Demanding e.g.~that the $U(1)_X$ charges allow the MSSM Yukawa couplings and the Weinberg operator, are flavor-blind and commute with $SU(5)$ for matter, we obtain the constraints
\begin{subequations}
\label{eq:SU5YukawaCouplings}
\begin{align}
2 q_{10} + q_{H_u} &= 2R  \,, \\
q_{10} + q_{\bar{5}} + q_{H_d} &= 2R \,, \\
2q_{\bar{5}} + 2q_{H_u} &= 2R \,,
\end{align}
\end{subequations}
where $R=0$ or $R=1$ distinguishes non-$R$ and $R$ symmetries. Assuming no light Higgs triplets, one thus finds for the anomalies
\begin{subequations}
\begin{align}
  A_{SU(3)^2-U(1)_X}  &= \frac32 \left(3 q_{10} + q_{\bar{5}}\right) -3R\,,\\ 
  A_{SU(2)^2-U(1)_X}  &= \left(3 q_{10} + q_{\bar{5}}\right) -3R\,,\\
  A_{U(1)_Y^2-U(1)_X} &= \frac35\left(3 q_{10} + q_{\bar{5}}\right) -\frac92R \,.
\end{align}
\end{subequations}
We use the conventions $\ell\!\left(\boldsymbol{N}\right)=\frac{1}{2}$ and
$\ell\!\left(\boldsymbol{\text{adj}}_f\right)=N$ for $SU(N)$. Furthermore, while generically $U(1)$
normalizations in a bottom-up approach are not fixed, we use the GUT normalization for the hypercharge
$U(1)_Y$. Clearly, imposing anomaly universality is a strong additional constraint. This constraint is neither
required for consistency (as was discussed \linebreak in Chapter~2), nor particularly motivated from grand unification:
Assuming an \linebreak $SU(5)\times U(1)_X$ theory at some high scale, after the breaking to the MSSM we might expect
that \linebreak $A_{SU(3)^2-U(1)_X}=A_{SU(2)^2-U(1)_X}$. However, this only holds 
\begin{itemize}
  \item for non-$R$ symmetries, since for $R$ symmetries, the gauginos contribute a non-universal factor $\ell\!\left(\boldsymbol{\text{adj}}_f\right)$,
  \item and before doublet-triplet splitting, as removing the triplets will change $A_{SU(3)^2-U(1)_X}$ but not $A_{SU(2)^2-U(1)_X}$ (unless $q_{H_u}+q_{H_d}-2R=0$, in which case the $\mu$ term is not forbidden).
\end{itemize}
From this it also follows that if the anomaly was universal before doublet--triplet splitting, it will not be afterwards and vice versa.

The situation is analogous for the case of a $\mathbbm Z_N$ instead of $U(1)_X$. In that case,
\eqref{eq:SU5YukawaCouplings} has to be satisfied only mod $N$ and the anomaly coefficients are only defined
mod $N$ or mod $\frac{N}{2}$, depending on whether $N$ is odd or even. 

\medskip
This discussion illustrates that anomaly universality leads to unnecessary constraints on potential
$U(1)_X$ or $\mathbbm Z_N$ symmetries. In particular, \eqref{eq:SU5YukawaCouplings} implies 
\begin{align}
2\left(q_{H_u} + q_{H_d}\right) = 10(R - q_{10}) \text{ mod } N\,,
\end{align}
and thus the $\mu$~term can be forbidden by both $R$ and non-$R$ symmetries once anomaly universality is not imposed.

\subsection{String model}

Now we want to construct an explicit string-derived realization illustrating the arguments made in Chapter~\ref{sec:GSMechanism}. We consider a model on the $T^6/\mathbbm Z_3$ orbifold in $E_8 \times E_8$ heterotic string theory. In the orbifold standard embedding with shift vector   
\begin{align}
  V = \frac13 \left(1, 1, -2, 0^5 \right) \left(0^8 \right)
  \label{eq:seshift}
\end{align}
and without Wilson lines, the gauge group is broken to $E_6 \times SU(3) \times E_8$. In particular, there is no anomalous $U(1)$. The orbifold action $\theta$ on the three two-tori with complex coordinates $z_i$ is defined as 
\begin{align}
   \theta \, : \, z_i \rightarrow e^{2\pi \I v_i}z_i \,, \quad v = (1,1,-2)\,,
\end{align}
and has 27 fixed points. The massless chiral spectrum is given by 
\begin{align}\label{eq:orbifoldspectrum}
  3\left(\boldsymbol{27}, \ol[.5]{\boldsymbol{3}}, \boldsymbol{1}\right)+
  27 \big[ \left(\boldsymbol{27}, \boldsymbol{1}, \boldsymbol{1}\right) +
  3 \left(\boldsymbol{1}, \boldsymbol{3}, \boldsymbol{1}\right)\big]\,.  
\end{align}
In order to construct a blowup, we replace the fixed points of the orbifold by 27 exceptional divisors $E_{\alpha \beta \gamma} \equiv E_r$, where $\alpha, \beta, \gamma =1\dots 3$ label the fixed points on the three tori and \mbox{$r = 1 \dots 27$}. The Abelian gauge flux, which is supported at the exceptional divisors only, is of the form $\mathcal F = E_r V_r^I H_I$, where $H_I$ denotes the elements of the Cartan subalgebra of $E_8 \times E_8$. The bundle vectors $V_r^I$ are chosen to coincide with shifted momenta of twisted orbifold states. In our case they read
\begin{subequations}
  \label{eq:bumodes}
  \begin{align}   
  V_1 &= \frac13 \left(~~\,2,~~2,~~2,~~\,0, 0^4 \right) \left(0^8 \right)\,,\\
  V_2 &= \frac13 \left(-1,-1,-1,~~3, 0^4 \right)\left(0^8 \right)\,, \\ 
  V_3 &= \frac13 \left(-1,-1,-1,-3, 0^4 \right)\left(0^8 \right)\,.
\end{align}
\end{subequations}
We assign $V_1$ to the first $k$ fixed points, $V_2$ to the next $p$ fixed points and $V_3$ to the remaining $q=27-k-p$ ones. The blowup procedure in the field theory picture corresponds to giving a vev to the twisted states with shifted momenta~(\ref{eq:bumodes}). For details on the blowup procedure, see e.g. \cite{Nibbelink:2007rd, Nibbelink:2008tv, Nibbelink:2009sp}. 
 
The bundle vectors have to satisfy three consistency conditions: The flux quantization condition, the Bianchi identity and the Donaldson--Uhlenbeck--Yau (DUY) equations. Since the bundle vectors are given by shifted momenta of the twisted states $\left(\boldsymbol{27}, \boldsymbol{1}, \boldsymbol{1}\right)$, the flux quantization condition $3V^I_r \in \Lambda_{E_8 \times E_8}$ is automatically fulfilled. It is also easy to check that the Bianchi identity $V_r^2 = \frac{4}{3}$ is fulfilled as well. The DUY equations read
\begin{align}
  \label{eq:duy}
  0=\frac{1}{2} \int_X J \wedge J \wedge \mathcal F = \frac{1}{2}\int_X J \wedge J \wedge E_r V_r^I H_I = \sum_r \text{vol}(E_r) V_r\,,
\end{align}
where $J=a_i R_i-b_r E_r$ denotes the K\"ahler form and $\text{vol}(E_r)>0$ in the K\"ahler cone $a\gg b>0$. Condition~(\ref{eq:duy}) can be fulfilled for all configurations $(k,p,q)$ with arbitrarily large exceptional divisor volumes: The DUY equations can be written as
\begin{align}
	\sum_{r=1}^k \text{vol}(E_r) \cdot V_1 + \! \sum_{r=k+1}^{k+p}\! \text{vol}(E_r) \cdot V_2 + \! \!   	\! \sum_{r=k+p+1}^{k+p+q} \! \! \! \text{vol}(E_r) \cdot V_3 = 0\,.
\end{align}
Since in our model the bundle vectors add up to zero, this simplifies to the condition
\begin{align}
   \sum_{r=1}^k \text{vol}(E_r) = \! \sum_{r=k+1}^{k+p} \! \text{vol}(E_r) = \! \! \! \sum_{r=k+p+1}^{k+p   +q} \! \! \! \text{vol}(E_r)\,.
\end{align}

\medskip

Each of the bundle vectors in Eq.~(\ref{eq:bumodes}) breaks $E_6$ to a differently embedded $ SO(10)\times U(1)$, but only two breakings are independent. We are thus left with the gauge group   
\begin{align}
  G = SO(8) \times U(1)_A \times U(1)_B \times SU(3) \times E_8\,.
  \label{eq:ggroup}
\end{align}
The two $U(1)$ factors are generated by
\begin{align}
  t_A = \left(2,2,2,0,0^4\right)\left(0^8\right)\,, \qquad t_B = \left(0,0,0,2,0^4\right)\left(0^8\right)\,.
  \label{eq:u1gens}
\end{align}
Using this normalization, all charges are integer. The $U(1)$ generators are related to the bundle vectors~(\ref{eq:bumodes}) via
\begin{align}\label{eq:genrelations}
  V_1 = \frac{1}{3} t_A\,, \quad V_2 = -\frac{1}{6} t_A + \frac{1}{2}t_B\,, \quad V_3 = - \frac{1}{6}t_A-\frac{1}{2}t_B\,. 
\end{align} 
The induced decomposition of the $\boldsymbol{27}$  (via $SO(10)\times U(1)$) is
\begin{align}
  \begin{split}
    \label{eq:branching}
    \mathbf{27} &\longrightarrow \mathbf{16}_{1} + \mathbf{10}_{-2} + \mathbf{1}_{4}\\
    &\longrightarrow
    (\mathbf{8}_\text{\scshape s})_{1,-1} + (\mathbf{8}_\text{\scshape c})_{1,1} +
    (\mathbf{8}_\text{\scshape v})_{-2,0} + \mathbf{1}_{-2,-2} + \mathbf{1}_{-2,2} +
    \mathbf{1}_{4,0}\,.     
  \end{split}
\end{align}
The 3 bundle vectors \eqref{eq:bumodes} on the Calabi--Yau side correspond on the orbifold side to a vev of the 3 singlets in \eqref{eq:branching}. The massless chiral spectrum on the blowup depends on the distribution $(k,p,q)$ of the bundle vectors over the 27 fixed points. The untwisted sector and the twisted $\left(\boldsymbol 1, \boldsymbol 3 \right)$'s in the orbifold spectrum~(\ref{eq:orbifoldspectrum}) always contribute $72 \cdot \left(\boldsymbol 1, \boldsymbol 3 \right)$ and $9 \cdot\left( \boldsymbol 8, \boldsymbol 3\right)$, while some of the twisted $\boldsymbol{8}$'s get massive or vector-like. As a result, we get 
\begin{align}
 (p-q) \cdot \left(\boldsymbol{8}_\text{\scshape v},\boldsymbol{1}\right)+ (k-q) \cdot \left(\boldsymbol{8}_\text{\scshape s},\boldsymbol{1}\right) + (k-p)\cdot \left(\boldsymbol{8}_\text{\scshape c},\boldsymbol{1}\right)\,.
\end{align}
Furthermore, the $U(1)$ charges of the states from the twisted sector will be shifted by the charges of the blowup modes at the respective fixed point. The blowup mode at the $r$-th fixed point is redefined as
\begin{align}\label{eq:expbumodes}
\Phi_r^\text{BU-Mode} = e^{b_r + i\beta_r}\,,
\end{align}
where $b_r$ are the K\"ahler (size) moduli of the $r$-th blowup cycle dual to $E_r$, and $\beta_r$ are the localized axions. In addition, the twisted orbifold states $\Phi^\text{Orb}$ get redefined as
\begin{align}\label{eq:buredefinition}
\Phi^\text{BU}_r =  \Phi^\text{Orb} \cdot e^{-(b_r+i\beta_r)}\,.
\end{align}
With this, the $U(1)$ charges of the blowup states are given by 
\begin{align}
q(\Phi^\text{BU}) = q(\Phi^\text{Orb}) - q(\Phi^\text{BU-Mode})\,.
\end{align}
For a detailed discussion of this redefinition procedure and a complete match of the orbifold and blowup spectra, see \cite{Blaszczyk:2011ig, GrootNibbelink:2007ew}.

\subsubsection{Anomaly coefficients}

Once the spectrum on the blowup is determined, we can calculate the anomaly coefficients of $U(1)_A$ and $U(1)_B$ via the corresponding triangle graphs using (\ref{eq:anomcoeffs}). As an important consistency check, the result can be compared with the coefficients appearing in the 4d anomaly polynomial~(\ref{eq:4dpolynomial}).

From the structure of the anomaly polynomial we draw a number of conclusions: First we take (\ref{eq:4dpolynomial}) for the first $E_8$ (and hence drop the primes on all field strengths) and express it in terms of the second Chern class $c_2$ as follows
\begin{align}
  I_6 = \frac{1}{(2\pi)^6} \int_X \left[
    \frac{1}{6}\tr\!\left(\mathcal{F}F\right)^2 - \frac{1}{4} c_2\tr F^2 +
    \frac{7}{48} c_2 \tr R^2 \right] \tr\!\left(\mathcal{F}F\right)\,. 
\end{align}
Here we have used that $\tr \mathcal{R}^2 = -2c_2$ and $\tr \mathcal{F}^2 = \tr \mathcal{R}^2$ in cohomology, as long as the second $E_8$ remains unbroken. We expand the flux and the 4d field strength as $\mathcal{F} = E_r V_r^I H_I$ and $F=F^I H_I$ and insert $k$ times the contribution from $V_1$, $p$ times the contribution from $V_2$ and $q$ times the contribution of $V_3$ to obtain
\begin{align}
	\begin{split}\label{eq:nicepolynomial}
	I_6 &\sim F_A^3 \cdot \left( \frac{k-6}{12} \right) + F_A F_B^2\cdot \left( \frac{k-18}{4} \right)\\ 
	&\quad + F_A\left[ \tr F_{SU(3)}^2 + \tr F_{SO(8)}^2 - \frac{7}{12} \tr R^2\right] \cdot \left(\frac{k-9}{2} 	\right) \\
	 &\quad+ F_B \left[ \frac{1}{8} F_B^2 + \frac{1}{48} F_A^2 + \tr F_{SU(3)}^2 + \tr F_{SO(8)}^2 - 	\frac{7}{12} \tr R^2 \right] \cdot \left( \frac{p-q}{2} \right)\,. \\
	\end{split}
\end{align}
We have also used relations~(\ref{eq:genrelations}) and $\int_X c_2\, E_r V_r^I F_I = -6 \sum_r V_r^I F_I $. From~(\ref{eq:nicepolynomial}) it is evident that $U(1)_B$ is omalous if $p=q$, that the cubic $U(1)_A$ anomaly vanishes for $k=6$, and that the non-Abelian anomalies of $U(1)_A$ vanish for $k=9$. In particular, there is no configuration with omalous $U(1)_A$. 

Although $U(1)_B$ is omalous in some bundle configurations, both $U(1)$'s are always massive. The axion associated with $U(1)_B$ always feels a shift proportional to $\lambda$ due to the transformation~(\ref{eq:deltaB2}) and hence gets a mass via the St\"uckelberg mechanism. We expand $B_2 = b_2 + \alpha_i R_i - \beta_r E_r$ to see from~(\ref{eq:deltaB2}) that $\beta_r$ transforms as 
\begin{align}
   \delta \beta_r = \tr \lambda V_r\,,
\end{align} 
while the $\alpha_i$ do not play a role in this case. Therefore the anomalies in our model are canceled by the non-universal axions $\beta_r$ only. There can be no contribution from the universal $b_2$ because of the absence of an anomalous $U(1)$ on the orbifold. In particular,~(\ref{eq:genrelations}) indicates that the first $k$ of the $\beta_r$ cancel the anomalies of $U(1)_A$ and the rest cancel a mixture of both $U(1)$'s. Hence the axions can be redefined such that only two of the $1+27$ possible axions transform with a shift, and the others are invariant under~(\ref{eq:anomalousvarition}). 

Another way to see that both $U(1)$'s are massive is by looking at the mass term for the 4d gauge bosons arising from the action~(\ref{eq:GSaction}) 
\begin{align}
\int_X H_3 \wedge \ast H_3 = A_\mu^I A^{\mu \,J} \left(M^2\right)^{IJ} + \dotsm \,,
\end{align}
with the mass matrix
\begin{align}\label{eq:mmatrix}
 \left(M^2\right)^{IJ} = V_r^I V_s^J \cdot \int_X E_r \wedge \ast_6 \,E_s\,.
\end{align}
Here, $\ast_6$ denotes the six-dimensional Hodge star. Using \cite{Strominger:1985ks}
\begin{align}
\ast_6 \, E_s = \frac34\,\frac{\text{vol}(E_s)}{\,\text{vol}(X)} J \wedge J - \frac{1}{2} E_s \wedge J
\end{align}
and the DUY equations \eqref{eq:duy}, we find for \eqref{eq:mmatrix}
\begin{align}
 \left(M^2\right)^{IJ} =  \frac{9}{2} V_r^I V_s^J \delta_{rst} b_t = \frac{1}{2} 
 \begin{pmatrix}
  m_1 & m_1 & m_1 & m_2 \\
  m_1 & m_1 & m_1 & m_2 \\
  m_1 & m_1 & m_1 & m_2 \\
  m_2 & m_2 & m_2 & m_3
\end{pmatrix}\, ,
\end{align}
and $\left(M^2\right)^{IJ} = 0$ for $I,J>4$. The entries $m_i$ are given by
\begin{align}
 \begin{split}
   m_1 &= 4\sum_{r=1}^k b_r   +  \sum_{r=k+1}^{k+p} b_r + \sum_{r=k+p+1}^{27} b_r\,,\qquad m_2 =-3\sum_{r=k+1}^{k+p} b_r + 3\sum_{r=k+p+1}^{27} b_r\,,\\
   m_3 &= 9\sum_{r=k+1}^{k+p} b_r + 9\sum_{r=k+p+1}^{27} b_r\,,
 \end{split}
\end{align}
which means that $(M^2)^{IJ}$ always has rank two in the blowup phase.

\begin{table}[t]
  \begin{center}
    \renewcommand\arraystretch{1.3}
   \subfloat[Chiral massless spectra of the three example models.]{\label{tab:spectra}
    \begin{tabular}{|c|p{7.2cm}|}
      \hline
      $(k,p,q)$ 	&	\hfill Massless chiral spectrum \hfill{} \\
      \hline
      \hline
       $(9,9,9)$	&	$24\left(\boldsymbol{1},\boldsymbol{3}\right)_{4,0} +					24\left(\boldsymbol{1},\boldsymbol{3}\right)_{-2,-2} +	
     	24\left(\boldsymbol{1}, \boldsymbol{3}\right)_{-2,2} \newline + 							3\left(\boldsymbol{8}_\text{\scshape v}, \boldsymbol{3}\right)_{-1,1} +
      	3\left(\boldsymbol{8}_\text{\scshape s}, \boldsymbol{3}\right)_{-1,-1} + 					3\left(\boldsymbol{8}_\text{\scshape c},\boldsymbol{3}\right)_{2,0}$	\\  
      \hline 
      $(25,1,1)$		&	$72\left(\boldsymbol{1}, \boldsymbol{3}\right)_{4,0} \newline +			24\left(\boldsymbol{8}_\text{\scshape s},\boldsymbol{1}\right)_{3,1} +
      	24\left(\boldsymbol{8}_\text{\scshape c},\boldsymbol{1}\right)_{3,-1} \newline + 			3\left(\boldsymbol{8}_\text{\scshape v},\boldsymbol{3}\right)_{-1,1} + 					3\left(\boldsymbol{8}_\text{\scshape s}, \boldsymbol{3}\right)_{-1,-1} +
      	3\left(\boldsymbol{8}_\text{\scshape c}, \boldsymbol{3}\right)_{2,0}$ \\  
      \hline
      $(13,13,1)$	&	$36\left(\boldsymbol{1}, \boldsymbol{3}\right)_{4,0} + 					36\left(\boldsymbol{1}, \boldsymbol{3}\right)_{-2,2}\newline +							12\left(\boldsymbol{8}_\text{\scshape v},\boldsymbol{1}\right)_{0,2} + 							12\left(\boldsymbol{8}_\text{\scshape s},\boldsymbol{1}\right)_{3,1}\newline + 				3\left(\boldsymbol{8}_\text{\scshape v}, \boldsymbol{3}\right)_{-1,1} + 					3\left(\boldsymbol{8}_\text{\scshape s},\boldsymbol{3}\right)_{-1,-1} + 					3\left(\boldsymbol{8}_\text{\scshape c}, \boldsymbol{3}\right)_{2,0}$ \\    
      \hline
    \end{tabular}}
    \renewcommand\arraystretch{1}
    
    \subfloat[Anomaly coefficients.]{ \label{tab:coeffs}
    \begin{tabular}{|c|c|c|c|c|c|c|}
      \hline
      $(k,p,q)$		&	$U(1)$ factor	&	$A_{SO(8)^2}$		&	$A_{SU(3)^2}$		&	$A_{U(1)_A^2}$		&
      $A_{U(1)_B^2}$	&	$A_\text{grav$^2$}$	\rule[-8pt]{0pt}{1pt}\\ 
      \hline
      \hline
      \multirow{2}{*}{$(9,9,9)$}	&	$U(1)_A$	&	 0	&	0	&	3888		&	-1296	&	0	\\ 
      \cline{2-7}
      &$U(1)_B$&	0	&	0	&	0	&	0	&	0	\\
      \hline
      \multirow{2}{*}{$(25,1,1)$}	&	$U(1)_A$	&	288	&	288	&	24624	&	1008	&	2016	 \\ 
      \cline{2-7}
      &$U(1)_B$&	0	&	0	&	0	&	0	&	0	\\
      \hline
      \multirow{2}{*}{$(13,13,1)$}	&	$U(1)_A$	&	72	&	72	&	9072	&	-720	&	504	 \\ 
      \cline{2-7}
      &$U(1)_B$&	72	&	72	&	1728	&	1728	&	504	\\      
      \hline
    \end{tabular}}
    \renewcommand\arraystretch{1}
 \caption{Details of the three example models. The gauge group is always $SO(8) \times U(1)_A \times U(1)_B \times SU(3)$, with the unbroken $E_8$ factor omitted.}
  \end{center}
\end{table}

\subsubsection{Examples}

As an example of the above statements we explicitly discuss three simple models with different $(k,p,q)$. Their spectra and anomaly coefficients are summarized in Table~\subref*{tab:spectra} and~\subref*{tab:coeffs}, respectively. Model 1 has the configuration $(k,p,q) = (9,9,9)$. As explained before, $U(1)_B$ is omalous and $U(1)_A$ only has Abelian anomalies. Model 2 has $(k,p,q) = (25,1,1)$. Here, $U(1)_B$ is still omalous and $U(1)_A$ has Abelian and non-Abelian anomalies. Model 3 with configuration $(k,p,q) = (13,13,1)$ is the most general, where all anomaly coefficients are nonzero.

In all models the anomaly coefficients from the triangle diagrams and from the anomaly polynomials match. All models exhibit Green--Schwarz anomaly cancellation with non-universal axions.

\section{Remnant discrete symmetries}
\label{sec:DiscreteSymmetries}

Discrete symmetries are very useful for explaining the absence of a perturbative $\mu$ term and of dimension-four and five proton decay operators. While discrete non-$R$ symmetries often stem from broken $U(1)$ symmetries, discrete $R$ symmetries arise as remnants of the internal Lorentz symmetry after compactification. 

\subsection[Non-$R$ symmetries]{Non-$\boldsymbol R$ symmetries}

Since the blowup is generated from twisted orbifold fields that get a vev, discrete symmetries can arise from remnants of the $U(1)$ gauge groups under which the blowup fields were charged. 

In the family of models at hand, the discrete non-$R$ symmetries are the ones left over from the two broken $U(1)$'s. One finds from the branching of the $\boldsymbol{27}$ of $E_6$ in \eqref{eq:branching} that the bundle vectors~(\ref{eq:bumodes}) correspond to the blowup modes  
\begin{align}
  (\boldsymbol{1}, \boldsymbol{1})_{4,0}\,, \quad (\boldsymbol{1}, \boldsymbol{1})_{-2,-2}\,, \quad
  (\boldsymbol{1}, \boldsymbol{1})_{-2,2}\,. 
\end{align}
When these get a vev, a $\mathbbm Z_2 \times \mathbbm Z_2$ subgroup survives\footnote{Generically, this is a $\mathbbm Z_4 \times \mathbbm Z_4$. In our family of models however, all states have charge 0 or 2 under both $\mathbbm Z_4$'s.}. It is easy to check that both $\mathbbm Z_2$ factors are omalous. 

\subsection[$R$ symmetries]{$\boldsymbol R$ symmetries}
The discussion of remnant $R$ symmetries is more involved. $R$ symmetries are those transformations that do not commute with supersymmetry, which in superspace language means that the Grassmann coordinate $\theta$ transforms nontrivially. Since there is only one such coordinate in 4d $\mathcal{N}=1$ supersymmetry, this can at most  be a single $U(1)$ or $\mathbbm{Z}_N$: If there are several such symmetries, they can be redefined such that only one of them transforms $\theta$, while the others act as usual non-$R$ symmetries. The normalization is commonly chosen such that $\theta$ transforms with charge~1, which implies that the superpotential $W$ has charge~2. This convention only fixes the charges of the fields up to an admixture of non-$R$ symmetries that leave $\theta$ invariant. Furthermore, a $\mathbbm{Z}_2$ $R$ symmetry can be turned into a non-$R$ symmetry by a combination with a sign reversal on the fermions, so $\mathbbm{Z}_2$ symmetries do not lead to true $R$ symmetries.

 In the following, we begin with reviewing $R$ symmetries from the orbifold point of view. After that, we discuss them from the Calabi--Yau perspective.  

\subsubsection[$R$ symmetries on the orbifold]{$\boldsymbol R$ symmetries on the orbifold} 
The orbifold $(T^2)^3/\mathbbm{Z}_3$ possesses a discrete $(\mathbbm{Z}_3)^3$ rotational symmetry stemming from rotating each torus independently by $\frac{2\pi \I}{3}$ (note that this is a symmetry of the compactification space but not an orbifold space group element). In the literature \cite{Kobayashi:2004ya}, one often finds the $R$ charge of the orbifold state defined as 
\begin{align}
 \label{eq:RCharges}
 R^i=q_{\text{sh}}^i-\widetilde{N}^i+\widetilde{N}^{*\,i}\,.
\end{align}
Here, $q_{\text{sh}}$ is the shifted $SO(8)$ momentum of the right-movers in light-cone gauge (sometimes called $H$-momentum), and the $\widetilde{N}^i$ and $\widetilde{N}^{*\,i}$ are integer oscillator numbers. The combination of the oscillators is such that the $R^i$ are invariant under picture changing. The $R$ charges of the space-time fermions are related to the above $R$ charges via $q_{\text{sh}}^\text{f}=q_{\text{sh}}-\left\{\frac12,\frac12,\frac12,\frac12\right\}$, and hence $R^i_\text{f}=R^i-\frac12$. $R$~charge conservation requires that for a superpotential coupling involving $L$ chiral superfields $\Phi_\alpha$, $W\supset \Phi_1\dotsm \Phi_L$, the charges satisfy
\begin{align}
 \label{eq:RconservationOrbifold}
 \sum_{\alpha=1}^L R_{\alpha}^i \equiv 1 \text{ mod } N_i\,,\quad i=1,2,3\,.
\end{align}
Here $N_i$ is the order of the orbifold twist in the $i^\text{th}$ torus. It should be noted that in this convention the superpotential $W$ has $R$  charge $1$ and thus  $\theta$ has $R$ charge $\frac12$. 

In cases where rotating a sub-torus independently by $\frac{2\pi \I }{N_i}$ is a symmetry, an orbifold state $\Phi$ transforms as 
\begin{align}
  \label{eq:Rtrafo}
  \mathcal{R}\,:~\Phi\to e^{2\pi \I v\cdot R}\,\Phi
\end{align}
with $v=\left(0,\frac{1}{N_1},0,0\right)$ and similarly for the other sublattice rotations. Explicitly, this transformation acts in the following way:
\begin{itemize}
  \item The bosonic $R$ charge from  (\ref{eq:RCharges}) is quantized in units of $\frac{1}{N_i}$, so under~(\ref{eq:Rtrafo}) bosons get a phase $e^{\frac{2\pi\I}{N_i^2}}\,$. 
  \item The $R$ charges of the fermions is shifted by $-\frac{1}{2}$, so $\theta$ transforms with a phase $e^{\frac{2\pi\I}{2N_i}}$, i.e.\ sublattice rotations act as a $\mathbbm{Z}_{2N_i}$ $R$~symmetry.
  \item Finally, the order of~(\ref{eq:Rtrafo}) acting on the fermions is given by the least common multiple of $N_i^2$ and $2N_i$.
\end{itemize}
To summarize, the $R$ transformations form a $\mathbbm{Z}_{2N_i^2}$ symmetry, under which the charges of bosons, fermions and $\theta$ are of the form $2k$, $2k-N$ and $N$, respectively, where $k$ is an integer. If $N$ is even, so are all charges, and consequently, only a $\mathbbm{Z}_{N_i^2}$ is realized on the fields.

\medskip

Owing to this slightly confusing symmetry pattern, one finds at least 3 different $R$ charge normalizations in the literature:  
\begin{enumerate}
 \item $W$ has charge~1, and the smallest charge quantization is in units of $\frac{1}{2N_i}$. This is inspired by the orbifold $R$~rule (\ref{eq:RconservationOrbifold}).
 \item $W$ has charge~2, and the smallest charge quantization is in units of $\frac{1}{N_i}$, which fits with the usual four-dimensional $R$ symmetry conventions.
 \item $W$ has charge $2 N_i$, and the smallest charge quantization is in units of $1$.
\end{enumerate}
In the case at hand, each 2-torus can be rotated independently (i.e.\ $N_i=3$), and we use the second normalization, such that we speak of a $\mathbbm{Z}_{6}^R$ symmetry where fermion charges are quantized in multiples of $\frac{1}{3}$, bosonic ones in multiples of $\frac23$, and $\theta$ has charge $1$. Note that in particular the twisted states $\Phi$ corresponding to the $\mathbf{27}$ of $E_6$ have $R=\frac13(0,1,1,1)$ and thus transform with charge $\frac19$ under each $\mathbbm{Z}_3$ sublattice rotation. Clearly, this is not a {\itshape bona fide} $\mathbbm{Z}_{6}$ symmetry because applying it six times does not give the identity of the fields, but it fits with the standard $R$ symmetry normalization from four-dimensional supersymmetry, and the orbifold $R$ charge conservation~(\ref{eq:RconservationOrbifold}) becomes a $\text{mod }6$ condition.

\subsubsection[$R$ symmetries from the blowup perspective]{$\boldsymbol R$ symmetries from the blowup perspective} 

To find unbroken $R$ symmetries after switching on vevs to generate the blowup, we seek combinations of the three sublattice rotations $\mathcal{R}_i$ and the two $U(1)$ generators $T_{A,B}$ which leave the blowup modes invariant, 
\begin{align}
  \boldsymbol{1}_{q_A,q_B}&\longrightarrow \left(\mathcal{R}_1\right)^r   \left(\mathcal{R}_2\right)^s   \left(\mathcal{R}_3\right)^t \left(T_A\right)^{q^A}
  \left(T_B\right)^{q^B} \boldsymbol{1}_{q_A,q_B} =\boldsymbol{1}_{q_A,q_B}\,,
\end{align}
for $\left(q_A,q_B\right)=(4,0)$, $(-2,2)$ and $(-2,-2)$. This implies that $r+s+t\equiv 0\text{ mod }3$, i.e.\ only a trivial $\mathbbm{Z}_2$ $R$ symmetry remains. Note that by combining with discrete non-$R$ $\mathbbm{Z}_N$ symmetries, higher $\mathbbm{Z}_N$ $R$ symmetries (with $N>3$) can be obtained. For the examples presented here, the discrete non-$R$ symmetries are $\mathbbm{Z}_2\times\mathbbm{Z}_2$, such that in this case no $R$ symmetry enhancement by mixing in other symmetries is possible. Hence for the models at hand it is expected that no nontrivial $R$ symmetry will be left after blowing up. However, as discussed in Section~\ref{sec:BottomUpModel}, both $R$ and non-$R$ symmetries can forbid the unwanted superpotential terms once anomaly universality is not required.

We will now investigate how to reproduce this from the perspective of the resolution space. One way to uncover discrete $R$ symmetries on the resolution Calabi--Yau manifolds is to look at the GLSM realization. For the sake of clarity, we focus on the $(k,p,q)=(9,9,9)$ model, where the blowup can be described with just three exceptional divisors. However, using the results from \cite{Blaszczyk:2011hs}, the analysis can be repeated for more general configurations and for other orbifolds in the same fashion. We leave the full analysis including more realistic cases for future work and merely outline the procedure. 
\begin{table}[t]
 \begin{center}
   \begin{tabular}{|l||ccc|ccc|ccc|ccc|ccc|ccc|}
     \hline
     Charges & $z_{11}$  & $z_{12}$ & $z_{13}$ & $z_{21}$ & $z_{22}$ & $z_{23}$ & $z_{31}$ & $z_{32}$ &
     $z_{33}$ & $x_{111}$ & $x_{211}$ & $x_{311}$ & $c_{1}$ & $c_{2}$ & $c_{3}$ \\ 
     \hline
     \hline
     $R_1$      &  1&1&1  &  0&0&0  &  0&0&0 &  0&0&0  & -3&0&0 \\
     $R_2$      &  0&0&0  &  1&1&1  &  0&0&0 &  0&0&0  & 0&-3&0 \\
     $R_3$      &  0&0&0  &  0&0&0  &  1&1&1 &  0&0&0  & 0&0&-3 \\
     $E_{111}$  &  1&0&0  &  1&0&0  &  1&0&0 &  -3&0&0 & 0&0&0 \\
     $E_{211}$  &  0&1&0  &  1&0&0  &  1&0&0 &  0&-3&0 & 0&0&0 \\
     $E_{311}$  &  0&0&1  &  1&0&0  &  1&0&0 &  0&0&-3 & 0&0&0 \\
     \hline
    \end{tabular}
  \caption{Charge assignment of the GLSM superfields describing the geometry in the notation of
    \cite{Blaszczyk:2011hs}.\label{tab:GLSMCharges}}
  \end{center}
\end{table}

The geometry is realized as the blowup of a complete intersection in $\left(\mathbbm{P}^2[3]\right)^3/\mathbbm{Z}_3$. The coordinates and charges for the GLSM realization are given in Table~\ref{tab:GLSMCharges}. The orbifold twist acts with a phase $e^{\frac{2\pi\I}{3}}$ on $z_{11}$, $z_{21}$ and $z_{31}$. From this and Table~\ref{tab:GLSMCharges}, one derives the $F$-term equations for the $c_i$ 
\begin{subequations}
  \label{eq:fterms}
  \begin{align}
    0&=z_{11}^3x_{111}+z_{12}^3x_{211}+z_{13}^3x_{311}\,,\\
    0&=z_{21}^3x_{111}x_{211}x_{311}+z_{22}^3+z_{23}^3\,,\\
    0&=z_{31}^3x_{111}x_{211}x_{311}+z_{32}^3+z_{33}^3\,,
  \end{align} 
\end{subequations}
specifying the complete intersection, and the $D$-term equations
\begin{subequations}
  \label{eq:dterms}
  \begin{align}
    |z_{1\alpha}|^2+|z_{21}|^2+|z_{31}|^2-3|x_{\alpha11}|^2 &=b_{\alpha11}\,, \qquad \alpha\in\{1,2,3\}\,,\\
    \sum_{\rho=1}^3|z_{i\rho}|^2 &=a_i\,,\qquad\quad\, i\in\{1,2,3\}\,,
  \end{align}
\end{subequations}
specifying the geometric phase (the vevs of the $c_i$ have already been set to zero). The $z_{i\rho}$ correspond to the inherited divisors, where $i=1,2,3$ labels the torus and $\rho=1,2,3$ labels the fixed point. The three coordinates $x_{\alpha11}$ label the three exceptional divisors where each resolves 9 of the 27 orbifold fixed points. Finally, the FI parameters $a_i$ and $b_{\alpha11}$ are related to the sizes of the tori and the exceptional divisors, respectively. We have chosen a phase where $a_i\gg0$ and $a_i\gg b_{\alpha11}$. For $b_{\alpha11}\gg0$, one uncovers the blowup regime, while $b_{\alpha11}\rightarrow-\infty$ yields the orbifold regime. (The notation has been chosen as in \cite{Blaszczyk:2011hs}.)

To find $R$ symmetries in this picture, we have to find holomorphic automorphisms of the ambient space which leave \eqref{eq:fterms} and~\eqref{eq:dterms} invariant, and under which the holomorphic $(3,0)$--form $\Omega$ transforms nontrivially \cite{Witten:1985xc}. This is true because $\Omega$ is related to the four-dimensional SUSY generators via the internal spinor $\eta$ by $\Omega_{ijk}=\eta^T\Gamma_{ijk}\eta$, where $\Gamma_{ijk}$ is a product of 3 gamma matrices. $\Omega$ can acquire at most a phase $\gamma=e^{2\pi \I \alpha},~\alpha\in\mathbbm{R}$, i.e.\ $\Omega\mapsto \gamma\Omega$. This means that $\eta\rightarrow \pm \gamma^{\frac12}$ and thus the superpotential $W$ transforms as $W \rightarrow \gamma W$, i.e.\ like $\Omega$. On the orbifold, the twisted $\boldsymbol{27}^3$ coupling is allowed, so the $\boldsymbol{27}$ of $E_6$ has to transform with a phase $\gamma^{\frac13}$. 

In our case, we find that the $F$- and $D$-term constraints are invariant under the $\mathbbm{Z}_3$ transformations\footnote{Note that not all of these symmetries are independent, since some can be related using the GLSM $U(1)$ charges.}
\begin{align}
   z_{i\alpha} \rightarrow e^{\frac{2\pi \I }{3}\cdot k_{i\alpha}} z_{i \alpha}\qquad\forall~i,\,\alpha\,.
\label{eq:Z3ZSymmetries}
\end{align}
Furthermore, there is the $\mathbbm Z_3$ symmetry
\begin{align}
   (x_{111},x_{211},x_{311})\rightarrow e^{\frac{2\pi \I }{3}\cdot k} (x_{111},x_{211},x_{311})\,.
\label{eq:Z3XSymmetries}
\end{align}
It should be noted that the presence of these symmetries is inherited from the symmetries of the orbifold. In other words, the polynomials in \eqref{eq:fterms} are not the most general ones in $(\mathbbm{P}^2[3])^3$ but have been chosen to be compatible with the orbifold action. In particular, the complex structure of the elliptic curves has been frozen at $\tau=e^{\frac{2\pi \I }{3}}$, so that we are already at a special sublocus of the whole moduli space which exhibits enhanced symmetries. At even more special points in moduli space, there appear certain symmetries under coordinate exchange: When $a_2=a_3$, there is a symmetry
\begin{align}\label{eq:GLSMpermutationa}
  z_{2\alpha}\leftrightarrow z_{3\alpha}\,, \quad \alpha=1,2,3\,.
\end{align}
When $b_1=b_2=b_3$, we find an $S_3$ permutation symmetry acting on
\begin{align}\label{eq:GLSMpermutationb}
  \left\{\left(z_{11},x_{111}\right),\left(z_{12},x_{211}\right),\left(z_{13},x_{311}\right)\right\}.
\end{align}
We can interpret these as exchanges of exceptional or inherited divisors, which are symmetries whenever the corresponding volumes, given by the K\"ahler parameters $a_i$ and $b_\alpha$, are equal. Focusing on the $\mathbbm{Z}_3$ symmetries, we find combinations such that $\Omega\rightarrow e^{\frac{2\pi \I }{3}}\Omega$. Thus $\gamma=e^{\frac{2\pi \I }{3}}$ and the $\boldsymbol{27}$ of $E_6$ transforms with $e^{\frac{2\pi \I }{9}}$, which reproduces the quantization in multiples of $\frac19$ from the orbifold.

So far, we have used the GLSM merely as a book-keeping device to realize the geometry of the blowup space, but it contains more information. In particular, from the preceding discussion it seems that the $\mathbbm{Z}_3$ symmetries~\eqref{eq:Z3XSymmetries} cannot be broken in the GLSM, since the $z_{i\alpha}$ appear only cubed or as absolute values. This seems puzzling, since all $R$ symmetries are generically broken from the orbifold point of view. On the other hand, from the GLSM point of view the $\mathbbm{Z}_3$ symmetries are merely accidental symmetries, and we would expect them to be broken by quantum effects. However, up to now we have not incorporated the gauge bundle into the GLSM description. 
\begin{table}[t]
  \centering
  \begin{tabular}{|c|c|}
    \hline
    Representation & Bundles\\\hline\hline
    $\left(\boldsymbol{1},\boldsymbol{3}\right)$ & $\mathcal{O}\!\left(0,0,0,4,-2,-2\right)\oplus\mathcal{O}\!\left(0,0,0,-2,4,-2\right)\oplus\mathcal{O}\!\left(0,0,0,-2,-2,4\right)$ \\
    \hline
    $\left(\boldsymbol{8},\boldsymbol{3}\right)_{\text{\scshape{v,s,c}}}$ & $\mathcal{O}\!\left(0,0,0,2,-1,-1\right)$, $\mathcal{O}\!\left(0,0,0,-1,2,-1\right)$, $\mathcal{O}\!\left(0,0,0,-1,-1,2\right)$ \\
    \hline
    $\left(\boldsymbol{8},\boldsymbol{1}\right)_\text{\scshape{v,s,c}}$ & $\mathcal{O}\!\left(0,0,0,0,-3,3\right)$, $\mathcal{O}\!\left(0,0,0,-3,0,3\right)$, $\mathcal{O}\!\left(0,0,0,-3,3,0\right)$ \\
    \hline
  \end{tabular}
  \caption{The bundles whose cohomology groups determine the chiral spectrum. The number of left-chiral representations in each case is given by $h^1(V)-h^2(V)$.\label{tab:cohoms}}
\end{table}
To make contact to the blowup model, we consider the line bundle $\mathcal{L}=\mathcal{O}\!\left(0,0,0,2,-1,-1\right)^3\oplus\mathcal{O}\!\left(0,0,0,0,3,-3\right)$ in analogy to the blowup modes~(\ref{eq:bumodes}). The chiral spectrum is then given by various line bundle cohomology groups (see Table~\ref{tab:cohoms}). Using \texttt{cohomCalg} \cite{Blumenhagen:2010pv, cohomCalg:Implementation} we can reproduce the chiral spectrum of the $(9,9,9)$ model in Table~\subref*{tab:spectra}, which is in turn consistent with the orbifold picture. 

The transformation of the states under the discrete symmetries can also be calculated via \texttt{cohomCalg}. Starting from the symmetries \eqref{eq:Z3ZSymmetries} and \eqref{eq:Z3XSymmetries}, which are given in terms of their actions on the GLSM coordinates, we have to determine how they act on the respective cohomologies of our bundle restricted to the Calabi--Yau hypersurface. A priori, it is not clear that the restriction of the symmetry to the Calabi--Yau can be lifted to the gauge bundle. A lift of the discrete symmetry to the gauge bundle which is consistent with the bundle projection and which preserves the group action is known as an equivariant structure \cite{Anderson:2012yf,Blumenhagen:2011xn}. It can be shown that for $\mathbbm{Z}_M$ symmetries all line bundles admit an equivariant structure. Given an equivariant structure, we have to check how the relevant bundle cohomologies transform. As in the case of the chiral spectrum, this is done by relating the gauge bundle on the Calabi--Yau to the gauge bundle of the ambient space via the Koszul resolution. The transformation of the matter states is then given in terms of the action of the symmetry on the global sections\footnote{If the bundle is not globally generated, one can twist it by an equivariant ample line bundle and check the transformation for the twisted bundle.} of the gauge bundle, which are given by polynomials in the homogeneous coordinates of the ambient space. 

\begin{table}[t]
  \centering
  \subfloat[Charges of the chiral-Fermi fields and of the polynomials arising as kinetic deformations.]
  {
    \label{tab:LambdaNa}
    \begin{tabular}{|c||c|c|c||c|c|c|c|c|c|c|c|}
      \hline 
      Charges &  $\Lambda^a$ & $\Lambda^4$ & $\Lambda^I$ &$N_{ab}$ & $N_4^a$ & $N^{a4}$ & $N_{aI}$ & $N_a^I$ & $N_{4I}$ & $N_4^I$ \\
      \hline 
      \hline 
      $E_{111}$ & 2  &  0 & 0 & -4 & 2 & 2 & -2 & -2 & 0 & 0 \\ 
      \hline
      $E_{211}$ & -1 &  3 & 0 & 2 & -4 & 2 & 1  & 1 &-3 & -3 \\ 
      \hline
      $E_{311}$ & -1 & -3 & 0 & 2 & 2 & -4 & 1 & 1 & 3 & 3 \\
      \hline
    \end{tabular}
  }
  \\
  \subfloat[Some monomials contributing to the chiral massless spectrum.]
  {
    \label{tab:LambdaNb}
    \begin{tabular}{|l||l|}
      \hline
      Polynomial & Some contributing monomials\\
      \hline
      \hline
      $N_{ab}$\rule[0mm]{0mm}{4.6mm} & $x_1^2\left(z_{21} \bar{z}_{22}\right)^2, \left(\bar{z}_{11}^2 z_{12}z_{13} \right)^2, x_1 \bar{x}_2 \bar{x}_3 \bar{z}_{21}z_{22}$\\
      \hline
      $N_{a,I}$, $N_a^I$\rule[0mm]{0mm}{3.5mm} & $x_2 x_3 z_{21}^2 \bar{z}_{22}^2, z_{11}^2 \bar{z}_{12}\bar{z}_{13}, \bar{x}_1 \bar{z}_{21} \bar{z}_{22}$\\
      \hline
      $N_{4I}$, $N_4^I$\rule[0mm]{0mm}{3.5mm}  & $\bar{x}_2 x_3, z_{12}^3 \bar{z}_{13}^3, x_1\bar{x}_3^2 z_{21}^3\bar{z}_{22}^3$\\ 
      \hline      
     \end{tabular}
  }
  \caption{Charges of the chiral-Fermi multiplets $\Lambda$ and the deformation coefficients $N$ and some of the contributing monomials. The monomials for $N_4^a$ and $N^{a4}$ can be obtained from $N_{ab}$ by permutations of indices.}  
\end{table}

Although the massless chiral spectra obtained from the orbifold, the blowup, and the GLSM perspective match, using the bundle $\mathcal{L}$ together with the geometric data of Table~\ref{tab:GLSMCharges} seems to lead to $U(1)$ anomalies in the GLSM. Recently, a mechanism has been proposed how these anomalies can be canceled \cite{Blaszczyk:2011ib,Quigley:2011pv}. A discussion of this is, however, beyond the scope of the paper.

Instead, we want to resort to the non-compact $\mathbbm C^3 /\mathbbm Z_3$ orbifold, where a consistent connection between the orbifold and the GLSM bundle description is known \cite{NibbelinkGroot:2010wm}. The bundle is described by chiral-Fermi multiplets $\Lambda^{\hat{I}}$, $\hat{I}=1,\dots,16$, which correspond to the Cartan subalgebra of $E_8 \times E_8$. The $\Lambda^{\hat{I}}$ are charged under the exceptional symmetries $E_{\alpha11}$, with charges given by the line bundle vectors \eqref{eq:bumodes} corresponding to the orbifold shifted momenta. Now the coordinates show up when determining the charged spectrum~\cite{NibbelinkGroot:2010wm}: The massless target space modes $\phi_{\text{4d}}\!\left(x^\mu\right)$ appear as deformations of the GLSM kinetic terms for the $\Lambda^{\hat{I}}$ as 
\begin{align}\label{eq:deformations}
  \int \D^2\theta^+ \phi_{\text{4d}} N^{\hat{J}}_{\hat{I}}\!\left(z_{i\alpha},x_{\alpha 11}\right)
  \Lambda^{\hat{I}} \ol[.5]{\Lambda}_{\hat{J}} + 
  \phi'_{\text{4d}} N_{\hat{I}\hat{J}}\!\left(z_{i\alpha},x_{\alpha 11}\right) \Lambda^{\hat{I}}
  \Lambda^{\hat{J}}  +\text{h.c.} 
\end{align}
Here the $N^{\hat{I}\hat{J}}$ and $N^{\hat{J}}_{\hat{I}}$ denote polynomials in the coordinate fields which are chosen such that the expression is gauge invariant. Note that this is a K\"ahler potential term, so the $N$'s need not be holomorphic. 

While locally at each fixed point the gauge group is $SU(3)\times SO(10)\times U(1)\times E_8$, the global model in the end has gauge group $SU(3)\times SO(8)\times U(1)^2\times E_8$. With regard to this, we split the index $\hat{I}$ into $\hat{I}=\left(a,4,I,\tilde{J}\right)$ with $a=1,2,3$, $I=5,\dots,8$. Furthermore, $\tilde{J}$ corresponds to the second $E_8$ which is unbroken and hence omitted in the following discussion. The gauge fields are determined by the neutral deformations $N_a^b$ and $N_4^4$ for $SU(3)\times U(1)^2$ and $N_I^J$ and $N_{IJ}$ for $SO(8)$. We can also read off the charged spectrum from the coefficients: $N_{ab}$, $N_4^a$ and $N^{a4}$ correspond to $\left(\boldsymbol{1},\ol[.5]{\boldsymbol{3}}\right)$ and $\left(\boldsymbol{1},\boldsymbol{3}\right)$, $N_{aI}$ and $N_a^I$ correspond to $\left(\boldsymbol{8},\ol[.5]{\boldsymbol{3}}\right)$, and $N_{4I}$ and $N_4^I$ correspond to $\left(\boldsymbol{8}, \boldsymbol{1}\right)$. The relevant charges of the bundle and the resulting polynomial charges are summarized in Table~\subref*{tab:LambdaNa}. Some of the contributing monomials are given in Table~\subref*{tab:LambdaNb}.  Note that the charges of the $N$'s reproduce some of the line bundle charges of Table~\ref{tab:cohoms}, but not all of them: The missing ones correspond to spinorial roots of $E_8$ which are not captured in the outlined procedure. Generically, the presence of the $N$'s in \eqref{eq:deformations} breaks at least some of the discussed $\mathbbm{Z}_3$ symmetries. However, a more thorough understanding of these deformations is needed, e.g.\ as to which monomials actually contribute in a given phase: Depending on the K\"ahler parameters, certain coordinates may or may not vanish, and this will play a role in determining the appearing operators, and hence the symmetry breaking. In particular, we should expect $R$ symmetries to reappear in the orbifold limit $b_{\alpha11}\to-\infty$.

\section{Conclusion} 
\label{sec:Conclusion}
Supersymmetry is an important ingredient in many extensions of the Standard Model. However, the MSSM in its simple form has a number of problematic operators, such as dimension-four and five operators violating baryon and lepton number conservation. Additionally, the $\mu$ term should be of the order of the weak or TeV~scale, rather than GUT or Planck scale. Many approaches to physics beyond the Standard Model address these problems by invoking additional symmetries, which forbid or suppress the unwanted terms. 

The aim of this paper is twofold. First, we discuss anomalies involving such symmetries and their cancellation. We point out that the Green--Schwarz mechanism does not require anomalies to be universal. We illustrate this point in two examples: First we discuss a bottom-up approach, where the MSSM is extended via an additional $U(1)$ or $\mathbbm{Z}_N$ symmetry. Then we give an explicit example of the heterotic string on blowups of the $T^6/\mathbbm{Z}_3$ orbifold. While there was no anomaly on the orbifold, there are two potentially anomalous $U(1)$'s in blowup. Depending on the particular blowup, they couple only among themselves or to the non-Abelian groups from the first $E_8$ and gravity (but never to the second $E_8$). 

Secondly, we analyze the appearance of (remnant) discrete symmetries. Our choice for the line bundles in blowup preserves a $\mathbbm{Z}_2\times\mathbbm{Z}_2$ subgroup. For $R$ symmetries, the situation is more involved: From the orbifold point of view, there are several conventions in the literature as to what the actual symmetry is, which we try to clear up. For the $\mathbbm{Z}_3$ orbifold, the sublattice rotations generate a $\mathbbm{Z}_{18}$ symmetry, with charge assignment
\begin{align}
  \left(\text{bosons},\text{fermions},\theta\right) &= \left(2k,2k-3,3\right)\,,\quad k\in\mathbbm{Z}\,.
\end{align}
Hence, the ``$R$ part'' of the symmetry is a $\mathbbm{Z}_6$. Our argument easily generalizes to other orbifolds, where the order of the sublattice rotations is the least common multiple of $N_i^2$ and $2N_i$. In the end, we find that the blowup generically breaks the $R$ symmetry down to a $\mathbbm{Z}_2$, which does not constitute a nontrivial $R$ symmetry. However, phenomenologically the presence of such $R$ symmetries is not necessarily required, as also non-$R$ symmetries can forbid the dangerous superpotential terms once anomaly universality is not imposed. We finally discuss a realization of the blowup in terms of a GLSM. Here we find indications that again the $R$ symmetries are generically broken by the gauge bundle, but at special subloci in K\"ahler moduli space, there appear additional symmetries which we interpret as exchange symmetries of divisors when the associated divisor volumes coincide. It would be interesting to see what remains of this effect when including Wilson lines.

\medskip

There are several issues which we defer to further work. On the phenomenological side, we should extend our analysis to more realistic models which could lead to the MSSM via (local) GUTs. From the theoretical point of view, we require a better understanding of the way the bundle breaks the $R$ symmetries in the GLSM, which is linked to the determination of the charged massless spectrum.
\subsection*{Acknowledgments}
We thank Michael Blaszczyk, Stefan Groot Nibbelink, Hans Jockers, Damian Kaloni Mayorga Pena and Hans Peter Nilles for useful discussions. This work was partially supported by the SFB-Transregio TR33 The Dark Universe (Deutsche Forschungsgemeinschaft), the European Union 7th network program Unification in the LHC era (PITN-GA-2009-237920) and the DFG cluster of excellence Origin and Structure of the Universe (Deutsche Forschungsgemeinschaft).


{\small
\providecommand{\href}[2]{#2}\begingroup\raggedright\endgroup
}


\begin{thebibliography}{10}

\bibitem{Gross:1984dd}
D.~J. Gross, J.~A. Harvey, E.~J. Martinec, and R.~Rohm ``{The Heterotic
  String}'' {\em Phys.Rev.Lett.} {\bf 54} (1985)
502--505.

\bibitem{Gross:1985fr}
D.~J. Gross, J.~A. Harvey, E.~J. Martinec, and R.~Rohm ``{Heterotic String
  Theory. 1. The Free Heterotic String}'' {\em Nucl.Phys.} {\bf B256} (1985)
253.

\bibitem{Green:1984sg}
M.~B. Green and J.~H. Schwarz ``{Anomaly Cancellation in Supersymmetric D=10
  Gauge Theory and Superstring Theory}'' {\em Phys.Lett.} {\bf B149} (1984)
  117--122.

\bibitem{Dixon:1985jw}
L.~J. Dixon, J.~A. Harvey, C.~Vafa, and E.~Witten ``{Strings on Orbifolds}''
  {\em Nucl.Phys.} {\bf B261} (1985)
678--686.

\bibitem{Dixon:1986jc}
L.~J. Dixon, J.~A. Harvey, C.~Vafa, and E.~Witten ``{Strings on Orbifolds.
  2.}'' {\em Nucl.Phys.} {\bf B274} (1986)
285--314.

\bibitem{Ibanez:1986tp}
L.~E. Ibanez, H.~P. Nilles, and F.~Quevedo ``{Orbifolds and Wilson Lines}''
  {\em Phys.Lett.} {\bf B187} (1987)
25--32.

\bibitem{Candelas:1985en}
P.~Candelas, G.~T. Horowitz, A.~Strominger, and E.~Witten ``{Vacuum
  Configurations for Superstrings}'' {\em Nucl.Phys.} {\bf B258} (1985)
46--74.

\bibitem{Aspinwall:1994ev}
P.~S. Aspinwall ``{Resolution of orbifold singularities in string theory}''
\href{http://www.arXiv.org/abs/hep-th/9403123}{[{\tt hep-th/9403123}]}.

\bibitem{Nibbelink:2007rd}
S.~Groot~Nibbelink, M.~Trapletti, and M.~Walter ``{Resolutions of $\mathbbm{C}^n/\mathbbm{Z}(n)$
  Orbifolds, their U(1) Bundles, and Applications to String Model Building}''
  {\em JHEP} {\bf 0703} (2007) 035
\href{http://www.arXiv.org/abs/hep-th/0701227}{[{\tt hep-th/0701227}]}.

\bibitem{Nibbelink:2007pn}
S.~Groot~Nibbelink, T.-W. Ha, and M.~Trapletti ``{Toric Resolutions of
  Heterotic Orbifolds}'' {\em Phys.Rev.} {\bf D77} (2008) 026002
\href{http://www.arXiv.org/abs/0707.1597}{[{\tt 0707.1597}]}.

\bibitem{NibbelinkGroot:2010wm}
S.~Groot~Nibbelink ``{Heterotic orbifold resolutions as (2,0) gauged linear
  sigma models}'' {\em Fortsch.Phys.} {\bf 59} (2011) 454--493
\href{http://www.arXiv.org/abs/1012.3350}{[{\tt 1012.3350}]}.

\bibitem{Blaszczyk:2011hs}
M.~Blaszczyk, S.~Groot~Nibbelink, and F.~Ruehle ``{Gauged Linear Sigma Models
  for toroidal orbifold resolutions}''
\href{http://www.arXiv.org/abs/1111.5852}{[{\tt 1111.5852}]}.

\bibitem{Schellekens:1986xh}
A.~Schellekens and N.~Warner ``{Anomalies, Characters and Strings}'' {\em
  Nucl.Phys.} {\bf B287} (1987)
317.

\bibitem{Blumenhagen:2005ga}
R.~Blumenhagen, G.~Honecker, and T.~Weigand ``{Loop-corrected compactifications
  of the heterotic string with line bundles}'' {\em JHEP} {\bf 0506} (2005) 020
\href{http://www.arXiv.org/abs/hep-th/0504232}{[{\tt hep-th/0504232}]}.

\bibitem{GrootNibbelink:2007ew}
S.~Groot~Nibbelink, H.~P. Nilles, and M.~Trapletti ``{Multiple anomalous U(1)s
  in heterotic blow-ups}'' {\em Phys.Lett.} {\bf B652} (2007) 124--127
\href{http://www.arXiv.org/abs/hep-th/0703211}{[{\tt hep-th/0703211}]}.

\bibitem{Blaszczyk:2011ig}
M.~Blaszczyk, N.~G. Cabo~Bizet, H.~P. Nilles, and F.~Ruehle ``{A perfect match
  of MSSM-like orbifold and resolution models via anomalies}'' {\em JHEP} {\bf
  1110} (2011) 117
\href{http://www.arXiv.org/abs/1108.0667}{[{\tt 1108.0667}]}.

\bibitem{Lukas:1999nh}
  A.~Lukas and K.~S.~Stelle,
  {``Heterotic anomaly cancellation in five-dimensions''}
  {\em JHEP} {\bf 0001} (2000) 010
  {\ttfamily [hep-th/9911156].}


\bibitem{Blaszczyk:2010db}
M.~Blaszczyk, S.~Groot~Nibbelink, F.~Ruehle, M.~Trapletti, and P.~K.
  Vaudrevange ``{Heterotic MSSM on a Resolved Orbifold}'' {\em JHEP} {\bf 1009}
  (2010) 065
\href{http://www.arXiv.org/abs/1007.0203}{[{\tt 1007.0203}]}.

\bibitem{Anderson:2011ns}
L.~B. Anderson, J.~Gray, A.~Lukas, and E.~Palti ``{Two Hundred Heterotic
  Standard Models on Smooth Calabi-Yau Threefolds}'' {\em Phys.Rev.} {\bf D84}
  (2011) 106005
\href{http://www.arXiv.org/abs/1106.4804}{[{\tt 1106.4804}]}.

\bibitem{Anderson:2012yf}
L.~B. Anderson, J.~Gray, A.~Lukas, and E.~Palti ``{Heterotic Line Bundle
  Standard Models}''
\href{http://www.arXiv.org/abs/1202.1757}{[{\tt 1202.1757}]}.

\bibitem{Lee:2010gv}
H.~M. Lee, S.~Raby, M.~Ratz, G.~G. Ross, R.~Schieren, {\em et al.} ``{A unique
  $\mathbbm{Z}_4^R$ symmetry for the MSSM}'' {\em Phys.Lett.} {\bf B694} (2011) 491--495
\href{http://www.arXiv.org/abs/1009.0905}{[{\tt 1009.0905}]}.

\bibitem{Scrucca:2004jn}
C.~A. Scrucca and M.~Serone ``{Anomalies in field theories with extra
  dimensions}'' {\em Int.J.Mod.Phys.} {\bf A19} (2004) 2579--2642
\href{http://www.arXiv.org/abs/hep-th/0403163}{[{\tt hep-th/0403163}]}.

\bibitem{Banks:1991xj}
T.~Banks and M.~Dine ``{Note on discrete gauge anomalies}'' {\em Phys.Rev.}
  {\bf D45} (1992) 1424--1427
  \href{http://www.arXiv.org/abs/hep-th/9109045}{[{\tt hep-th/9109045}]}.
  Revised version.

\bibitem{Fujikawa:1979ay}
K.~Fujikawa ``{Path Integral Measure for Gauge Invariant Fermion Theories}''
  {\em Phys.Rev.Lett.} {\bf 42} (1979)
1195.

\bibitem{Fujikawa:1980eg}
K.~Fujikawa ``{Path Integral for Gauge Theories with Fermions}'' {\em
  Phys.Rev.} {\bf D21} (1980)
2848.

\bibitem{AlvarezGaume:1984dr}
L.~Alvarez-Gaume and P.~H. Ginsparg ``{The Structure of Gauge and Gravitational
  Anomalies}'' {\em Annals Phys.} {\bf 161} (1985)
423.

\bibitem{Adams:2010zy}
A.~Adams, O.~DeWolfe, and W.~Taylor ``{String universality in ten dimensions}''
  {\em Phys.Rev.Lett.} {\bf 105} (2010) 071601
\href{http://www.arXiv.org/abs/1006.1352}{[{\tt 1006.1352}]}.

\bibitem{Gmeiner:2002es}
F.~Gmeiner, S.~Groot~Nibbelink, H.~P. Nilles, M.~Olechowski, and M.~Walter
  ``{Localized anomalies in heterotic orbifolds}'' {\em Nucl.Phys.} {\bf B648}
  (2003) 35--68
\href{http://www.arXiv.org/abs/hep-th/0208146}{[{\tt hep-th/0208146}]}.

\bibitem{Buchmuller:2007qf}
W.~Buchm{\"u}ller, C.~L{\"u}deling, and J.~Schmidt ``{Local SU(5) Unification
  from the Heterotic String}'' {\em JHEP} {\bf 0709} (2007) 113
\href{http://www.arXiv.org/abs/0707.1651}{[{\tt 0707.1651}]}.

\bibitem{Nibbelink:2009sp}
S.~Groot~Nibbelink, J.~Held, F.~Ruehle, M.~Trapletti, and P.~K. Vaudrevange
  ``{Heterotic $\mathbbm{Z}$(6-II) MSSM Orbifolds in Blowup}'' {\em JHEP} {\bf 0903} (2009)
  005
\href{http://www.arXiv.org/abs/0901.3059}{[{\tt 0901.3059}]}.

\bibitem{Nibbelink:2008tv}
S.~Groot~Nibbelink, D.~Klevers, F.~Ploger, M.~Trapletti, and P.~K. Vaudrevange
  ``{Compact heterotic orbifolds in blow-up}'' {\em JHEP} {\bf 0804} (2008) 060
\href{http://www.arXiv.org/abs/0802.2809}{[{\tt 0802.2809}]}.

\bibitem{Strominger:1985ks}
A.~Strominger ``{Yukawa Couplings in Superstring Compactification}'' {\em
  Phys.Rev.Lett.} {\bf 55} (1985)
2547.

\bibitem{Kobayashi:2004ya}
T.~Kobayashi, S.~Raby, and R.-J. Zhang ``{Searching for realistic 4d string
  models with a Pati-Salam symmetry: Orbifold grand unified theories from
  heterotic string compactification on a $\mathbbm{Z}$(6) orbifold}'' {\em Nucl.Phys.} {\bf
  B704} (2005) 3--55
\href{http://www.arXiv.org/abs/hep-ph/0409098}{[{\tt hep-ph/0409098}]}.

\bibitem{Witten:1985xc}
E.~Witten ``{Symmetry Breaking Patterns in Superstring Models}'' {\em
  Nucl.Phys.} {\bf B258} (1985)
75.

\bibitem{Blumenhagen:2010pv}
R.~Blumenhagen, B.~Jurke, T.~Rahn, and H.~Roschy ``{Cohomology of Line Bundles:
  A Computational Algorithm}'' {\em J. Math. Phys.} {\bf 51} (2010) 103525
  \href{http://www.arXiv.org/abs/1003.5217}{[{\tt 1003.5217}]}.

\bibitem{cohomCalg:Implementation}
``{\texttt{cohomCalg} package}.'' Download link 2010.
\newblock High-performance line bundle cohomology computation based on
  \cite{Blumenhagen:2010pv}.
  http://wwwth.mppmu.mpg.de/members/blumenha/cohomcalg.

\bibitem{Blumenhagen:2011xn}
R.~Blumenhagen, B.~Jurke, and T.~Rahn ``{Computational Tools for Cohomology of
  Toric Varieties}'' {\em Adv.High Energy Phys.} {\bf 2011} (2011) 152749
\href{http://www.arXiv.org/abs/1104.1187}{[{\tt 1104.1187}]}.

\bibitem{Blaszczyk:2011ib}
M.~Blaszczyk, S.~Nibbelink~Groot, and F.~Ruehle ``{Green-Schwarz Mechanism in
  Heterotic (2,0) Gauged Linear Sigma Models: Torsion and NS5 Branes}'' {\em
  JHEP} {\bf 1108} (2011) 083
\href{http://www.arXiv.org/abs/1107.0320}{[{\tt 1107.0320}]}.

\bibitem{Quigley:2011pv}
C.~Quigley and S.~Sethi ``{Linear Sigma Models with Torsion}'' {\em JHEP} {\bf
  1111} (2011) 034
\href{http://www.arXiv.org/abs/1107.0714}{[{\tt 1107.0714}]}.

\end{thebibliography}
\end{document}